\documentstyle[prl,aps,multicol]{revtex}

\newcommand{\bleq}{\ifpreprintsty
                   \else
                   \end{multicols}\vspace*{-3.5ex}{\tiny
                   \noindent\begin{tabular}[t]{c|}
                   \parbox{0.493\hsize}{~} \\ \hline \end{tabular}}
                   \fi}
\newcommand{\eleq}{\ifpreprintsty
                   \else
                   {\tiny\hspace*{\fill}\begin{tabular}[t]{|c}\hline
                    \parbox{0.49\hsize}{~} \\
                    \end{tabular}}\vspace*{-2.5ex}\begin{multicols}{2}
                    \fi}
\newcommand{\bcols}{\ifpreprintsty\else\begin{multicols}{2}\fi}
\newcommand{\ecols}{\ifpreprintsty\else\end{multicols}\fi}

\begin{document}
\title{Electromagnetic Field Correlation inside a Sonoluminescing Bubble}
\author{Pritiraj Mohanty}
\address{Condensed Matter Physics 114-36, California Institute of Technology, 
Pasadena, CA 91125}
\date{\today}
\maketitle

\begin{abstract}

We consider the correlation of the electromagnetic field to determine 
spatial coherence inside a sonoluminescing bubble. We explicitly calculate the first 
order correlation function for two limiting cases of the excitation field: a blackbody 
spectrum and a discrete multifrequency spectrum. The correlation length for blackbody 
fields at temperatures between 3000 K and 10000 K is found to be on the order of the 
optical wavelength, increasing with decreasing temperatures. We predict spectral lines 
in the emission spectrum of single bubble sonoluminescence in cooler bubbles with 
interior temperatures below 10000 K.

\vskip 0.2in

\end{abstract}

\bcols

Sonoluminescence is the outburst of a short light pulse from a noble gas 
bubble acoustically levitated in a liquid by ultrasound \cite{gaitan-crum,revs-son90s}. 
Some of its salient features are the short 
duration of outburst with a pulsewidth---typically on the order of 
picoseconds\cite{gompf,hiller,moran}, the large number of photons per outburst---typically on the order 
of a million, and the wide emission spectrum over and probably beyond the visible range. 
Recently, it has been argued that cooperative decay of excited states of metastable 
molecules could explain the light emission so peculiar in 
SL\cite{mohanty,mohanty2,brodsky}. 

The theory of collective decay or lasing\cite{mohanty,mohanty2} of the 
population-inverted medium of molecules naturally contains the short timescale 
of the sonoluminescence pulse and the large number of photons emitted per outburst. 
In single bubble sonoluminescence, the emitted spectrum, however, 
is found to be a continuum, often devoid of any structure resembling the spectral lines 
of the gas molecules of either the solute or the solvent. It is natural then to assume,
though naively, that the emission mechanism is not compatible with a lasing process
which is expected to be monochromatic. This argument is not correct for the 
reason that cooperative emission does indeed occur for a collection of molecules with a 
broad distribution of wavelengths. The input driving field which establishes this 
polychromatic distribution can be correlated over a length  comparable to the 
relevant scale in sonoluminescence.
This scale defines the length over which molecules are correlated 
through the common electromagnetic field. It also determines the number of
correlated molecules $N_{eff}$ and hence the number of emitted photons. 
A large $N_{eff}$ is essential in the proposed 
mechanism\cite{mohanty,mohanty2} as it determines the coherent part of the
decay, $I_{coherent} \propto N_{eff}^2$; the incoherent part of the emission goes as 
$N$, the total number of emitting entities.

In this paper we address this specific question pertaining to the excitation of 
the gas molecules necessary for population inversion. 
Considering that the gas in the sonoluminescing bubbles can 
reach a temperature of 10000 K ($\sim$ 1 eV), or even higher \cite{moss,lohse}, 
it is important to determine if there is correlation in the electric field at these 
temperatures. Towards that end,
we calculate the first order correlation function for a generalized 
excitation field in two limits of high ($k_BT \gg \hbar\omega$) 
and low ($k_BT \ll \hbar\omega$) interior temperature where the excitation
is dominated by (a) a blackbody field and (b) a discrete
multifrequency field arising from atomic/molecular lines, respectively.  
For the blackbody field, we find that the first-order correlation 
length in the electric field is on the order of the wavelength in the visible 
range, which allows the collection of excited molecules to be correlated over this 
length scale. With increasing temperature this correlation length is found to
decrease substantially.
Discrete multiple frequency fields corresponding to the spectral lines 
are also found to have long correlation lengths. 
We find that the correlation in the 
excitation is dominated by a blackbody field in hottor bubbles 
whereas discrete multiple frequency fields dominate at temperatures 
below 10000 K. 
 
We predict that the 
spectral lines will be observable in single bubble sonoluminescence (SBSL) 
at temperatures below 10000 K; these discrete lines are
expected to be prominent in the blue part of the emission spectrum.
The discrete lines were not  observed in 
the SBSL spectrum in previous experiments because of extremely high 
temperatures inside the bubbles.
This explains the experimental observation \cite{matula} 
that the sodium D-lines  are visible in multiple bubble sonoluminescence 
(MBSL), which are now known to be at a 
typical gas temperature of 5000 K\cite{suslick}; the
D-lines were absent in the SBSL spectrum because the single bubbles were much hotter. 
Our analysis allows us to understand, for the first time,  the connection between
multiple bubble sonoluminescence  \cite{suslick,arakeri} and single bubble 
sonoluminescence, namely they occur due to the same mechanism though in different 
temperature regimes.

Coherence in an atom-field system is best understood by studying 
the electromagnetic field modes\cite{glauber,mehta,born-wolf}. 
If $N$ excited atoms distributed over a range of frequencies decay 
coherently, the photon number
in the corresponding modes of the number state of the electromagnetic
field coupled to the atoms will change accordingly. 
Since the atoms and the field
form a closed system, it is then equivalent and even useful 
to consider coherence in the field modes. 

\par

The degree of coherence is often
expressed by the cross correlation 
$\langle E({\bf r}_1,t_1)E({\bf r}_2,t_2)\rangle$ and higher order correlations
$\langle E({\bf r}_1,t_1)E({\bf r}_2,t_2).....\rangle$
in the field variable $E({\bf r},t)$.  For example, the first order correlation is 
relevant for double-slit type interference experiments. The intensity is governed 
by the superposition of the amplitude contributions: 
$E_Q(t_2-t_1) = c_1 E({\bf r}_1,t_1) + c_2 E({\bf r}_2,t_2)$. This results in an 
additional part $2 Re\langle c_1 E({\bf r}_1,t_1) c_2 E({\bf r}_2,t_2) \rangle$ 
to the intensity $|E_Q(t_2-t_1)|^2$ coming from first order interference. 
This interference term 
is a complex function in general and it describes
the first order coherence (two-point correlation function
in the field). Similarly, higher order coherences are defined
by the higher order correlation functions. For example, four-point correlation
function defines the second order coherence relevant for
the Hanbury-Brown and Twiss experiment. Our calculation of higher order coherence 
which signifies purely quantum mechanical effects will be presented elsewhere. 

\par

%
%

The first order coherence function is defined by
\begin{equation}
C({\bf r}_1,{\bf r}_2;t_1,t_2) \equiv \langle E_i({\bf r}_1,t_1)E_j({\bf r}_2,t_2) \rangle,
\end{equation}
where  $\langle \dots \rangle$ denotes the ensemble averaging
through a density matrix opertaor $\rho$ such that
$\langle E_iE_j\rangle = Tr[\rho E_iE_j]$, and $\rho = (1/Z)exp(-\beta H)$.
The definition of $\rho$ is physically important, and may include the
active medium transitions.
A multifrequency electric field  can be expanded in terms
of the monochromatic fields
$E_i({\bf r},t) = \sum_\omega E_{i\omega}({\bf r},t)$.
If the temperature inside the bubble is very high, 
then the dominant contribution to the correlation
function will come from the blackbody field, and hence, $\rho$
could be determined for a known temperature $T$. At intermediate
temperatures, the active medium becomes important and the
spectral contribution from the transitions in the given system
must be taken into account as well. In both cases $\rho$ factorizes
into operators for each frequency $\omega$
and $\rho = \Pi_\omega \rho_\omega$.
Each $\omega$ component of the field may be assumed to be independent
of each other, in which case the number state representation of the
field reduces to product of the number states of individual
frequencies, i.e. $|n_{\omega_1} n_{\omega_2} \dots\rangle = \Pi_{\omega_i}
|n_{\omega_i}\rangle $.
Then the correlation function has the following expression:

\begin{eqnarray}
\nonumber
\langle E_i({\bf r}_1,t_1) E_j({\bf r}_2,t_2)\rangle &=& \\ 
\sum_\omega \sum_{n_\omega}
\langle n_\omega|E_{i\omega}({\bf r}_1,t_1)E_{j\omega}({\bf r}_2,t_2)|n_\omega\rangle
\langle n_\omega|\rho_\omega|n_\omega\rangle .
\end{eqnarray}

\par

The product of fields at two points may be given as

\begin{eqnarray}
\nonumber
E_i({\bf r}_1,t_1)E_j({\bf r}_2,t_2) = \\ 
\sum_\omega {\omega^2 \over c^2}
        (a_\omega a_\omega^\dagger e^{-i\omega(t_1-t_2)} A_{i\omega}({\bf r}_1)
A_{i\omega}^
\dagger({\bf r}_2) + h.c. ),
\end{eqnarray}

\noindent
which follows from the definition of $E({\bf r},t) = {i\over c} \sum_\omega
\omega(a_\omega exp[-i\omega t] A_\omega(r)+ h.c.)$ assuming Coulomb gauge
in the interaction picture. The diagonal terms in 
$a_\omega$ and $a_\omega^\dagger$
such as $a_\omega a_\omega$ 
are ignored in the following.

\par
The quantization for each mode $\omega$ requires that 
$a_\omega a_\omega^\dagger
+a_\omega^\dagger a_\omega = {\hbar \over 2\omega} [ 2 n_\omega + 1]$. 
Ignoring the second term, which represents vacuum fluctuations, 
one obtains an expression for the symmetrized correlation function:

\bleq
\begin{eqnarray}
Tr[\rho(E_iE_j + E_jE_i)/2] 
= [\sum_{n_\omega=0}^\infty exp[-{n_\omega \hbar\omega \over kT}]]^{-1} 
\sum_\omega \sum_{n_\omega} {\hbar\omega \over c^2}
n_\omega exp[- {n_\omega\hbar\omega \over kT}] Re (exp[-i\omega(t_i-t_j)]
A_{i\omega}({\bf r}_1) A_{j\omega}^\ast({\bf r}_2)).
\end{eqnarray}
\eleq

In the above equation, for equal times, the summation over the occupation
number $n_\omega$ yields the usual Planck's law. 
We consider
the effect of active medium via a distribution function:
\begin{equation}
f(\omega) = g_{\omega_p} \delta(\omega - \omega_p),
\end{equation}
\noindent
where the strengths of different lines are given by $g_{\omega_p}$ and
$\omega_c$ is a cutoff frequency.
This function will depend on the composition of the gas (solute) inside 
the bubble and the
surrounding liquid (solvent). The effect of temperature and pressure inside
the bubble  
can be taken into account by replacing the $\delta$-function by a
Lorentzian with a full width at half maximum corresponding to 
inhomogeneous broadening.
$f(w,p)$ may also be a continuous
function in case of an ionic spectrum which are realized in very
hot bubbles. 
The final spectrum of emitted light would contain
both $f(\omega,p)$ and the Planck spectrum.
At high temperatures, the Planck distribution would dominate over the
emission line distribution. For the emission lines to be
observable in SBSL, the temperature inside the bubble must satisfy
the condition
\begin{equation}
g_{\omega_p}\hbar\omega_p \gg {\hbar\omega_p \over e^{\hbar\omega_p/k_BT} -1}.
\end{equation}
\noindent
At high temperatures, the right hand side of the inequality becomes
equal to $k_BT$. In order to see the lines in the emission spectrum of
SBSL at a wavelength $\lambda$, the interior temperature has to be reduced
below $T^\ast = hc/\lambda k_B$. The visible range of 200 nm $\rightarrow$ 700 nm in 
wavelength corresponds to a temperature range of 71000 K $\rightarrow$ 20000 K.

\par

Eq. 4 gives the complete expression of correlation of fields in any arbitrary
shape. For $r_i = r_j$ and $t_i = t_j$, it reduces to
the Planck distribution. The vector potential ${\bf A}(r)$ satisfies  
the Helmholtz equation:
\begin{equation}
\nabla^2 {\bf A}_\omega + {\omega^2 \epsilon(r)} {\bf A}_\omega = 0,
\end{equation}
\noindent
with appropriate boundary
conditions. For a gas bubble with an average dielectric constant $\epsilon_1$
immersed in a liquid with a dielectric constant $\epsilon_2$.
The solution of the Helmholtz equation is normally constructed from the
linear combination of the solutions of the
scalar field equation,
requiring that
${\bf E}_\perp$, $\epsilon{E_r}$, $B_r$ and ${\bf B}_\perp/\mu$
be continuous across the bubble surface, i.e. at $r=a$. With the use of
the boundary conditions, the
transverse electric (TE) and transverse magnetic (TM) mode solutions
$(A_r,A_\theta,A_\phi)^{TE,TM}$ can be constructed inside and outside
the bubble \cite{milton,eberlein}.  

\par

{\it I. Blackbody field:}
Analytic evaluation of the coherence function
with the exact solutions of Eq. (7) is rather complicated. Here we give results
for the limiting case of  wavelengths short 
compared to the minimum bubble radius, $r \gg \lambda$. 
However, in standard SBSL experiments 
the minimum bubble radius is usually on the order of the wavelength. 
Even though $r \simeq \lambda$, closed-form expressions
are possible only in this case. This condition is strictly
satisfied in the case of  millimeter-sized laser-induced 
bubbles\cite{gary}. 

Let us choose a direction for the wave vector $k$ as $(\sin\theta \cos\phi,
\sin\theta \sin\phi, \cos\theta)$. The coherence function in the 
longitudinal\cite{long} direction is given by

\begin{eqnarray}
\nonumber
C_1^{long}(r,t) \simeq {1 \over r^2} \int_{0}^{\infty}
[g_{\omega_p}\delta(\omega - \omega_p)\hbar\omega \\  
+ {\hbar\omega \over {e^{\hbar\omega/k_BT}
 - 1}} ] \hbar\omega d\omega [{\sin kr \over kr}
- \cos kr] \cos \omega\tau.
\end{eqnarray}

\noindent
where $\omega = kc$. If the thermal contribution dominates, $T \gg \hbar\omega/k_B$, 
then the spectral distribution is essentially Planckian. This high temperature
result was first obtained by Bourret\cite{bourret} who derived the second-order 
electric correlation tensor of blackbody radiation using techniques similar to
those employed in the theory of isotropic turbulence of an incompressible
fluid.  The high temperature limit of the above result\cite{mehta2} reduces to
\begin{eqnarray}
\nonumber
C_1^{long}(r,0) \sim {1 \over r^2} \int_0^{\infty} {k dk \over {e^{\alpha k}-1}}
({\sin kr \over kr} - \cos kr) \\ 
\propto {1 \over r^3}(1 -r {\partial \over \partial r}) {1 \over \alpha} L({r\over \alpha})
\end{eqnarray} 
\noindent
where,  $\alpha = c \tau_\beta$, and $\tau_\beta = \hbar/k_BT$ is the thermal
time. $L(x)= \pi/2 \coth{\pi x} - 1/2x$ is the standard Langevin function. 
The expression in Eq.~9 is 
plotted in Fig. 1 for three different temperatures of the blackbody field.
The scale over which correlation reduces to a characteristic value, the
coherence length $R_\phi$, 
is seen to vary between 0.5 to 0.2 $\mu m$ for bubble temperatures ranging
from 3000 K to 10000 K. This mesoscopic scale of coherence length is
\begin{figure}
 \vbox to 7cm {\vss\hbox to 6cm
 {\hss\
   {\includegraphics{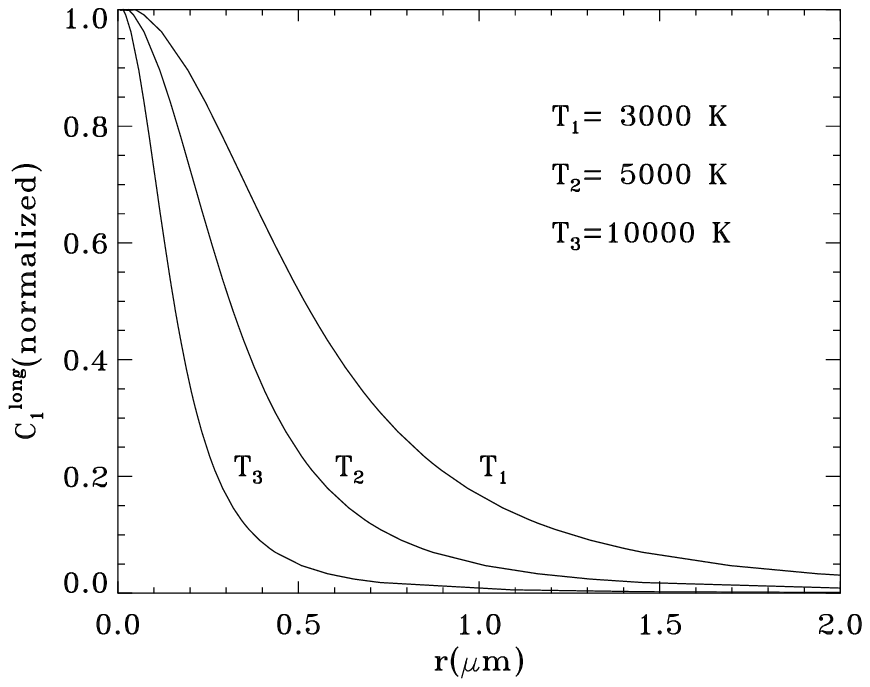}
   }
  \hss}
 }
\end{figure}
\noindent
FIG.1 Variation of the normalized coherence function at three different
temperatures.
\vskip 0.2in
\noindent
comparable to the compressed bubble size. At temperatures above 20000 K, the 
correlation length is found to be less than 0.1 $\mu m$.

{\it II. Discrete field:}
At low enough temperatures $g\hbar\omega_p \gg k_\beta T$, 
coherence in the field of a set of discrete emission lines can be
studied in a similar manner\cite{niez}. The coherence function 
$C_1^{long}$ in the long wavelength limit
becomes equal to  
\begin{equation}
C_1^{long} (k_p,r,0) \sim g_{\omega_p} k_p^3 [{1\over 3} -  
{(k_p r)^2 \over 30} + O((k_p r)^4)].
\end{equation}

The $r$ dependence of $C_1^{long}$ is parabolic, and a coherence 
length $R_\phi$ can be defined as the scale over which $C_1^{long}$ 
is reduced to a characteristic value,  half of its maximum. 
For a given wavelength
$\lambda_p$, 

\begin{equation}
R_\phi \sim \lambda_p/\pi.
\end{equation} 
\noindent

In the lowest order, coherence is seen to be strongly dependent on the 
wavelength $(\lambda_p = 2\pi/k_p)$, 
\begin{equation}
C_1^{long} \sim \lambda_p^{-3}. 
\end{equation}
Hence SL from the discrete 
spectral lines in the blue part of the spectrum should be stronger than 
that in the red part. This is true for the coherent radiation in the SL 
emission spectrum. Observation of such a trend is a strong indication
that the emitted light is indeed partly coherent.

In the blackbody regime, the correlation length is much less than
0.05 $\mu m$ at a typical temperature of 40000 K; thus the effective number of molecules
participating in collective decay reduces greatly as 
$N_{eff} = \rho (4\pi R_\phi^3/3)$, where $\rho$ is the average density. 
The sonoluminescence emission spectrum is not dominated by coherent
emission. $I_{incoherent} \sim N \propto R_{min}^3$, and $I_{coherent} 
\sim N_{eff}^2 \propto R_{\phi}^6$.  For a typical minimum bubble radius $R_{min}$ 
of 0.5 $\mu m$ and a total $N$ of $10^6$, the incoherent emission is comparable 
to the coherent emission. At higher temperatures, incoherent emission dominates.

Another corroborating evidence towards the transition from high temperature
to low temperature behavior can be obtained from the temporal shape of the 
emitted light pulse. The pulse shape should progressively change from either 
$sech^2$ or oscillatory behavior (signifying ringing) \cite{mohanty2} 
to exponential decay which signifies incoherent emission \cite{moss,lohse,frommhold}.

Let us summarize the 
three important aspects of the emission spectrum following our analysis: 
(a) Bubbles with an interior temperature much above 70000 K will typically 
display a blackbody emission spectrum in the entire visible range. 
As the temperature is reduced, spectral structure
will appear in the blue part of the emission spectrum, progressing towards
the red part. Experimentally, it is found that the spectrum is blackbody
and almost identical at long wavelengths, while it is gas specific at short
wavelengths \cite{hiller}, consistent with our picture. 
Below 10000 K, the emission spectrum
will show atomic/molecular lines in the entire visible range.
(b) The lines in the spectrum would be less strong in the red part as the
correlation length for a discrete field decreases with increasing
wavelength very sharply: $C_1^{long} \sim \lambda^{-3}$. 
(c) Our analysis provides a natural connection between SBSL and MBSL surrounding
the question as to why spectral lines are conspicuously absent in conventional 
SBSL experiments, but distinctly present in MBSL \cite{matula,suslick}. This is 
primarily because the gas temperature in MBSL is on the order of 5000 K, whereas in SBSL 
it's much higher.

In conclusion, we show that the electromagnetic field 
inside a sonoluminescing bubble has a finite correlation even at high temperatures. 
At high temperatures ($ T \gg$ 40000 K)
the emission spectrum is expected to be predominantly blackbody.
At low temperatures ($ T <$ 10000 K) discrete emission lines are predicted in single 
bubble sonoluminescence. Our analysis  explains the parametric
differences between multibubble sonoluminescence 
and single bubble sonoluminescence, exemplified by their very different emission spectra. 

I thank Prof. Lawrence Crum whose questions 
motivated this work. I am also grateful to Prof. Gary Williams, 
Ohan Baghdasarian, and Prof. Woowon Kang for helpful conversations.

\noindent

\ecols
\end{document}